\documentclass{PoS}

\title{Revised absolute amplitude calibration of the LOPES experiment}

\ShortTitle{LOPES - Revised amplitude calibration}

\author{
K.~Link$^{1}$,
\speaker{T.~Huege}$^{2}$,
W.D.~Apel$^{2}$,
J.C.~Arteaga-Vel\'azquez$^{3}$,
L.~B\"ahren$^{4}$,
K.~Bekk$^{2}$,
M.~Bertaina$^{5}$,
P.L.~Biermann$^{5,1}$,
J.~Bl\"umer$^{1,6}$,
H.~Bozdog$^{2}$,
I.M.~Brancus$^{7}$,
E.~Cantoni$^{4,8}$,
A.~Chiavassa$^{5}$,
K.~Daumiller$^{2}$,
V.~de~Souza$^{9}$,
F.~Di~Pierro$^{5}$,
P.~Doll$^{2}$,
R.~Engel$^{2}$,
H.~Falcke$^{10,3,5}$,
B.~Fuchs$^{1}$,
H.~Gemmeke$^{11}$,
C.~Grupen$^{12}$,
A.~Haungs$^{2}$,
D.~Heck$^{2}$,
R.~Hiller$^{2}$,
J.R.~H\"orandel$^{10}$,
A.~Horneffer$^{6}$,
D.~Huber$^{1}$,
P.G.~Isar$^{13}$,
K-H.~Kampert$^{14}$,
D.~Kang$^{1}$,
O.~Kr\"omer$^{11}$,
J.~Kuijpers$^{10}$,
P.~{\L}uczak$^{15}$,
M.~Ludwig$^{1}$,,
H.J.~Mathes$^{2}$,
M.~Melissas$^{1}$,
C.~Morello$^{8}$,
J.~Oehlschl\"ager$^{2}$,
N.~Palmieri$^{1}$,
T.~Pierog$^{2}$,
J.~Rautenberg$^{14}$,
H.~Rebel$^{2}$,
M.~Roth$^{2}$,
C.~R\"uhle$^{11}$,
A.~Saftoiu$^{7}$,
H.~Schieler$^{2}$,
A.~Schmidt$^{11}$,
S.~Schoo$^{2}$,
F.G.~Schr\"oder$^{2}$,
O.~Sima$^{16}$,
G.~Toma$^{7}$,
G.C.~Trinchero$^{8}$,
A.~Weindl$^{2}$,
J.~Wochele$^{2}$,
J.~Zabierowski$^{15}$,
J.A.~Zensus$^{5}$ - 
LOPES Collaboration \\
\llap{$^{1}$} Institut f\"ur Experimentelle Kernphysik, Karlsruher Institut f\"ur Technologie (KIT), Germany\\
\llap{$^{2}$} Institut f\"ur Kernphysik, Karlsruher Institut f\"ur Technologie (KIT), Germany\\
\llap{$^{3}$} Instituto de F\'isica y Matem\'aticas, Universidad Michoacana, Morelia, Mexico\\
\llap{$^{4}$} ASTRON, Dwingeloo, The Netherlands\\
\llap{$^{5}$} Dipartimento di Fisica, Universit\`a degli Studi di Torino, Torino, Italy\\
\llap{$^{6}$} Max-Planck-Institut f\"ur Radioastronomie, Bonn, Germany\\
\llap{$^{7}$} National Institute of Physics and Nuclear Engineering, Bucharest-Magurele, Romania\\
\llap{$^{8}$} Osservatorio Astrofisico di Torino, INAF Torino, Italy\\
\llap{$^{9}$} Universidade S$\tilde{a}$o Paulo, Instituto de F\'{\i}sica de S$\tilde{a}$o Carlos, S$\tilde{a}$o Carlos, Brasil\\
\llap{$^{10}$} Department of Astrophysics, Radboud University Nijmegen, The Netherlands\\
\llap{$^{11}$} Institut f\"ur Prozessdatenverarbeitung und Elektronik, KIT, Germany\\
\llap{$^{12}$} Faculty of Natural Sciences and Engineering, Universit\"at Siegen, Germany\\
\llap{$^{13}$} Institute for Space Sciences, Bucharest-Magurele, Romania\\
\llap{$^{14}$} Fachbereich C, Physik, Universit\"at Wuppertal, Germany\\
\llap{$^{15}$} Department of Astrophysics, National Centre for Nuclear Research, {\L}\'{o}d\'{z}, Poland\\
\llap{$^{16}$} Department of Physics, University of Bucharest, Bucharest, Romania\\

E-mail: \email{tim.huege@kit.edu}       
}
               
\abstract{}

\FullConference{The 34th International Cosmic Ray Conference,\\
		30 July- 6 August, 2015\\
		The Hague, The Netherlands}

\begin{document}

\section*{Abstract}
One of the main aims of the LOPES experiment was the 
evaluation of the absolute amplitude of the radio signal of air 
showers. This is of special interest since the radio technique offers 
the possibility for an independent and highly precise determination of 
the energy scale of cosmic rays on the basis of signal predictions from 
Monte Carlo simulations. For the calibration of the amplitude measured by 
LOPES we used an external source. Previous comparisons of LOPES 
measurements and simulations of the radio signal amplitude predicted by 
CoREAS revealed a discrepancy of the order of a factor of two. A 
re-measurement of the reference calibration source, now performed for 
the free field, was recently performed by the manufacturer. The updated
calibration values lead to a lowering of the reconstructed electric field
measured by LOPES by a factor of $2.6 \pm 0.2$ and therefore to a significantly better agreement with
CoREAS simulations. We discuss the updated calibration and its impact
on the LOPES analysis results.
\clearpage

\section{Introduction}

In the past decade, tremendous progress was made in the understanding 
of radio emission from extensive air showers and the reconstruction of 
cosmic ray properties from radio measurements \cite{huegereview}. One of the 
most challenging aspects of radio detection of extensive air showers 
turned out to be the absolute calibration of the detectors. In the 
historical experiments, there were in fact strong disagreements between 
results from different experiments, likely due to calibration issues 
\cite{atrashkevich}.

In LOPES \cite{FalckeNature2005}, we have made considerable effort from the very beginning 
of the experiment to provide a high-quality absolute calibration 
\cite{nehls}. Furthermore, we compared our measurements with predictions from Monte 
Carlo simulations with REAS3.11 \cite{reas} and CoREAS \cite{coreas}. These simulation 
codes calculate the radio emission from an extensive air shower from 
first principles (movement of charged particles plus classical 
electrodynamics using the endpoint formalism \cite{endpoints}). Given a 
set of parameters for the primary cosmic ray, the Monte Carlo 
simulation of the extensive air shower fully determines the result. 
The simulations contain no free parameters that could be tuned and thus 
provide an absolute prediction of the expected radio amplitude. This is very important, because such 
simulations can thus be used to calibrate the energy scale of cosmic ray detectors.

In a previous analysis \cite{simcomparison} we concluded that LOPES data were in agreement 
with predictions from the (simplified and thus obsolete) REAS3.11 simulation 
code, but were discrepant with predictions of the (more advanced and thus more precise)
CoREAS simulation code. The measured amplitudes were approximately a factor of 
two larger than those predicted by CoREAS. 
We have investigated possible causes for this discrepancy and 
have ruled out problems in the analysis procedure and the CoREAS simulation code.
It turned out, however, that the calibration of the LOPES antennas 
using an external reference source was based on unsuited calibration 
data for the reference source. In the following we describe how the calibration was updated, then 
illustrate the effects of the updated calibration on parameters derived from a data set of cosmic 
ray showers measured with LOPES, and finally present an updated 
comparison of LOPES data with CoREAS simulations.

\begin{figure}[b]
\centering
\includegraphics[angle=270,width=0.65\textwidth]{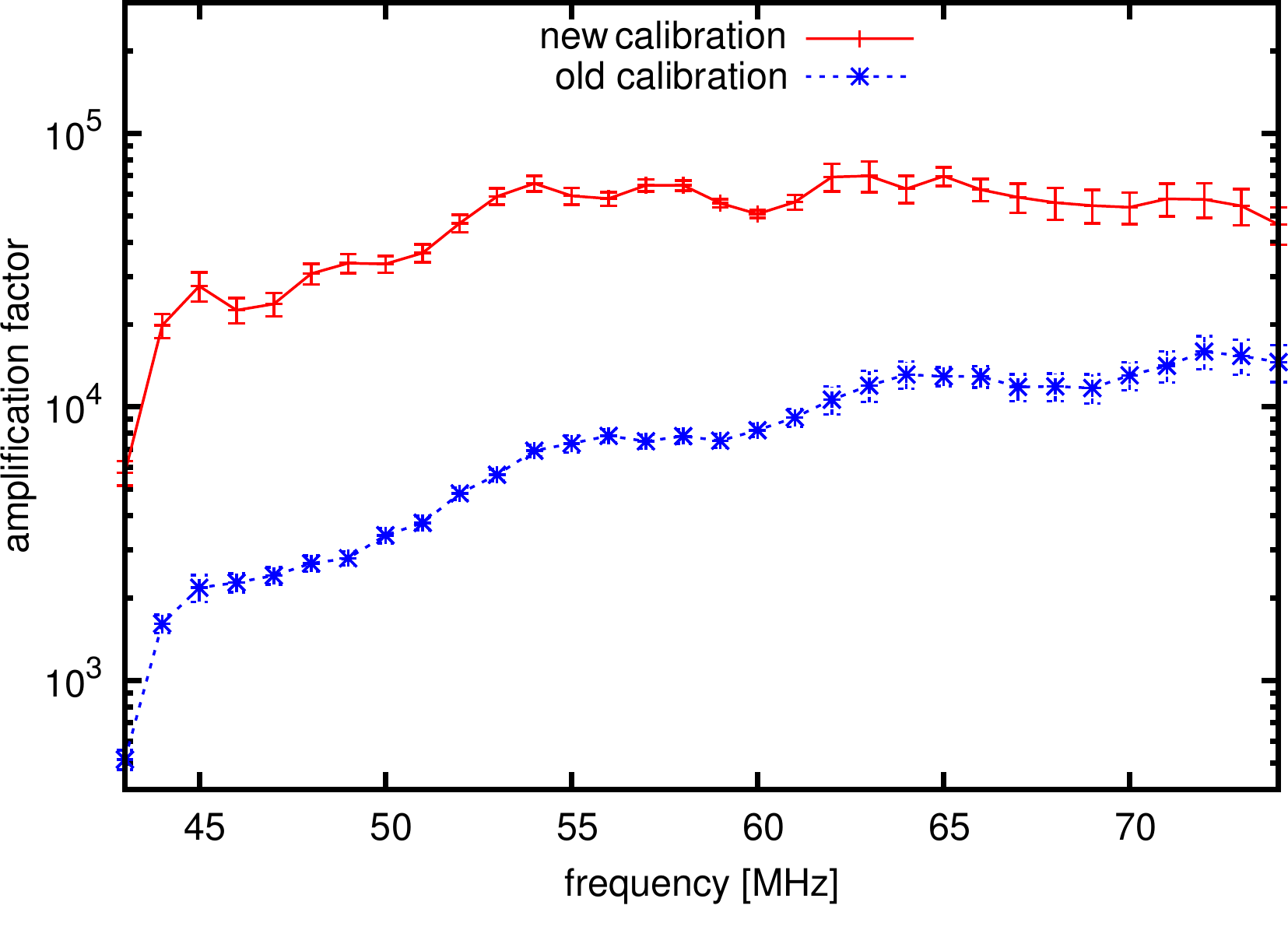}
\caption{Comparison of the amplification factors (related to power) of the analog chain 
of a LOPES antenna based on old and revised calibration data for the 
reference source.\label{fig:caliboldvsnew} The effect is the same for 
all antennas.}
\end{figure}

\section{Revised Amplitude Calibration}

The calibration of LOPES uses an external reference source 
\cite{nehls}, for which the manufacturer provided the absolute power radiated as a 
function of frequency. We deployed the calibration source 10~m above the
LOPES antennas, measured the power received in the 
LOPES antennas as a function of frequency, and then derived the amplification
factors of our analog chain as a function of frequency \cite{nehls}.

We performed an independent re-implementation of the analysis part of the 
calibration procedure, finding no problem that could explain the 
observed discrepancies between LOPES data and CoREAS simulations. Hence,
we contacted the manufacturer of the calibration source to verify the validity
of the provided calibration data. In detailed discussions it was realized that the 
calibration data we had originally received were acquired in a 
horizontal measurement setup with a reflective ground (\emph{free-field} 
conditions). However, the measurement of air showers by LOPES is better 
described with \emph{free-space} conditions. Therefore, we requested a re-calibration of the reference 
source and received updated calibration values from the manufacturer 
for \emph{free-space} conditions.

In Fig.\ \ref{fig:caliboldvsnew} we demonstrate the difference between 
the amplification factors for our analog chain as derived using the old
and new calibration data. The new values yield amplification factors that are 
systematically higher than those derived with the original calibration data. The same correction applies to all antennas
of the LOPES array, as the same reference source was used to calibrate 
all LOPES antennas. There is, however, a slight dependence of the 
correction between old and new calibration on frequency. We thus investigate in a detailed study 
how the updated calibration values influence the LOPES event 
reconstruction. 
We note that the systematic uncertainty on the absolute calibration of 
the LOPES amplitude scale previously reported as 35\% constituted the 
two-sigma uncertainty provided by the manufacturer of the reference 
source. The one-sigma uncertainty for the revised absolute amplitude calibration
amounts to 16\%.

\begin{figure*}[t]
    \centering
    \includegraphics[width=0.48\textwidth]{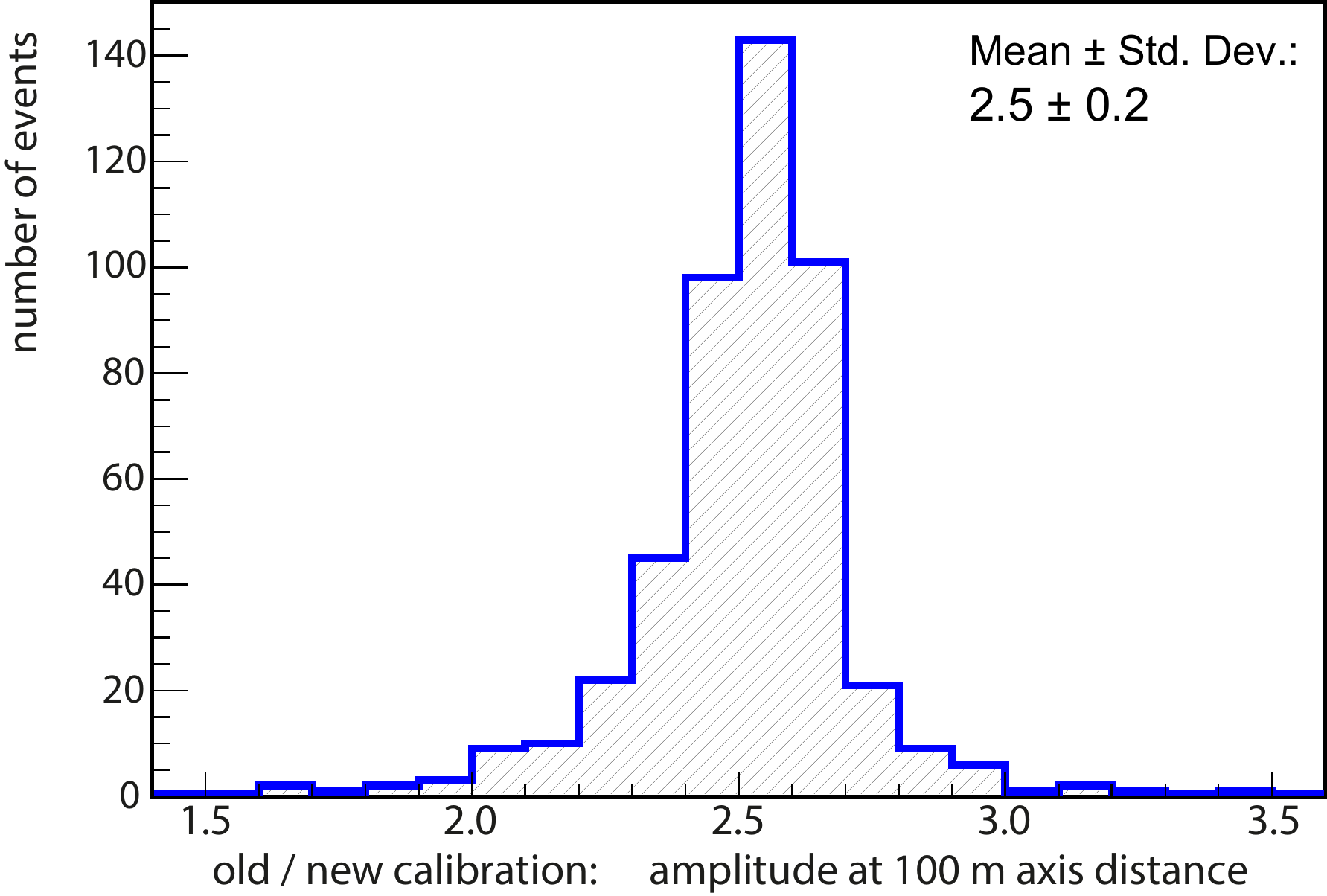}
    \includegraphics[width=0.48\textwidth]{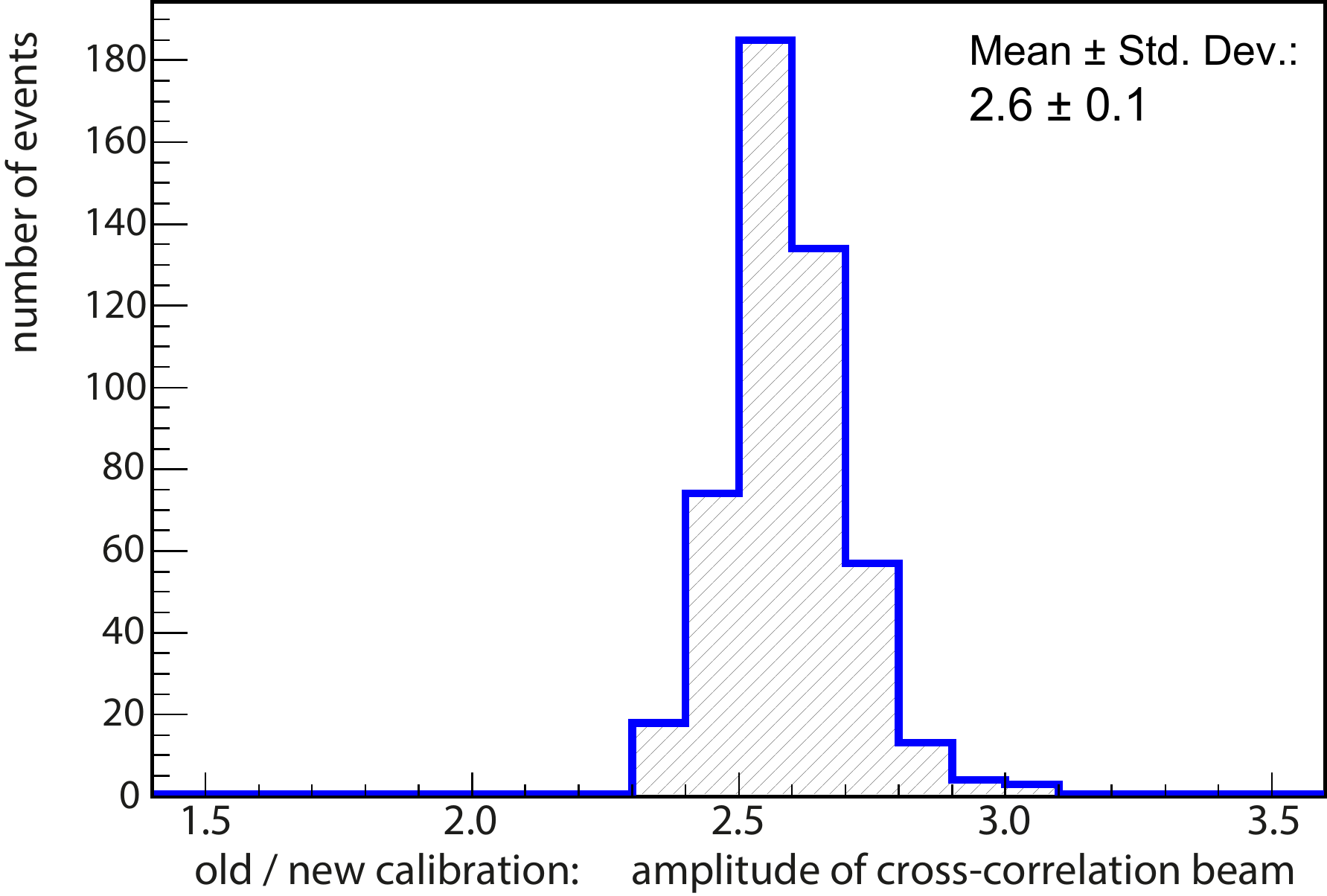}
    \caption{Influence of the revised calibration on the amplitude at 
    $100\,$m axis distance $\epsilon_{100}$ (left) and on the amplitude 
    of the cross-correlation beam (right) as measured with LOPES.}
    \label{fig:influenceofcalib}
\end{figure*}

\section{Influence of the new calibration}

We analyzed LOPES data acquired between the end of 2005 and the end of 
2009 with an energy reconstructed from KASCADE-Grande 
\cite{kascade,grande} data above $10^{17}\,$eV, a zenith angle below 
45$^{\circ}$, shower core inside the KASCADE or Grande fiducial areas,
and standard KASCADE-Grande cuts. Thunderstorm events are excluded.
The events are required to have a clear radio signal which is selected
via the cross-correlation beam, i.e., the correlated power must be larger
than 80\% of the full power, and the signal-to-noise ratio must be larger 
than 14, normalized with a factor $\sqrt{N_{ant}/30}$ to take into 
account the number of antennas available for a particular measurement.
With these conditions approximately 500 events remain for the 
analysis. In addition to the revised calibration, we made small improvements in our analysis, including 
an update of the antenna model to more realistic ground conditions, and 
an improved fitting algorithm for the lateral distribution function.



\begin{figure*}[t]
    \centering
    \includegraphics[width=0.48\textwidth]{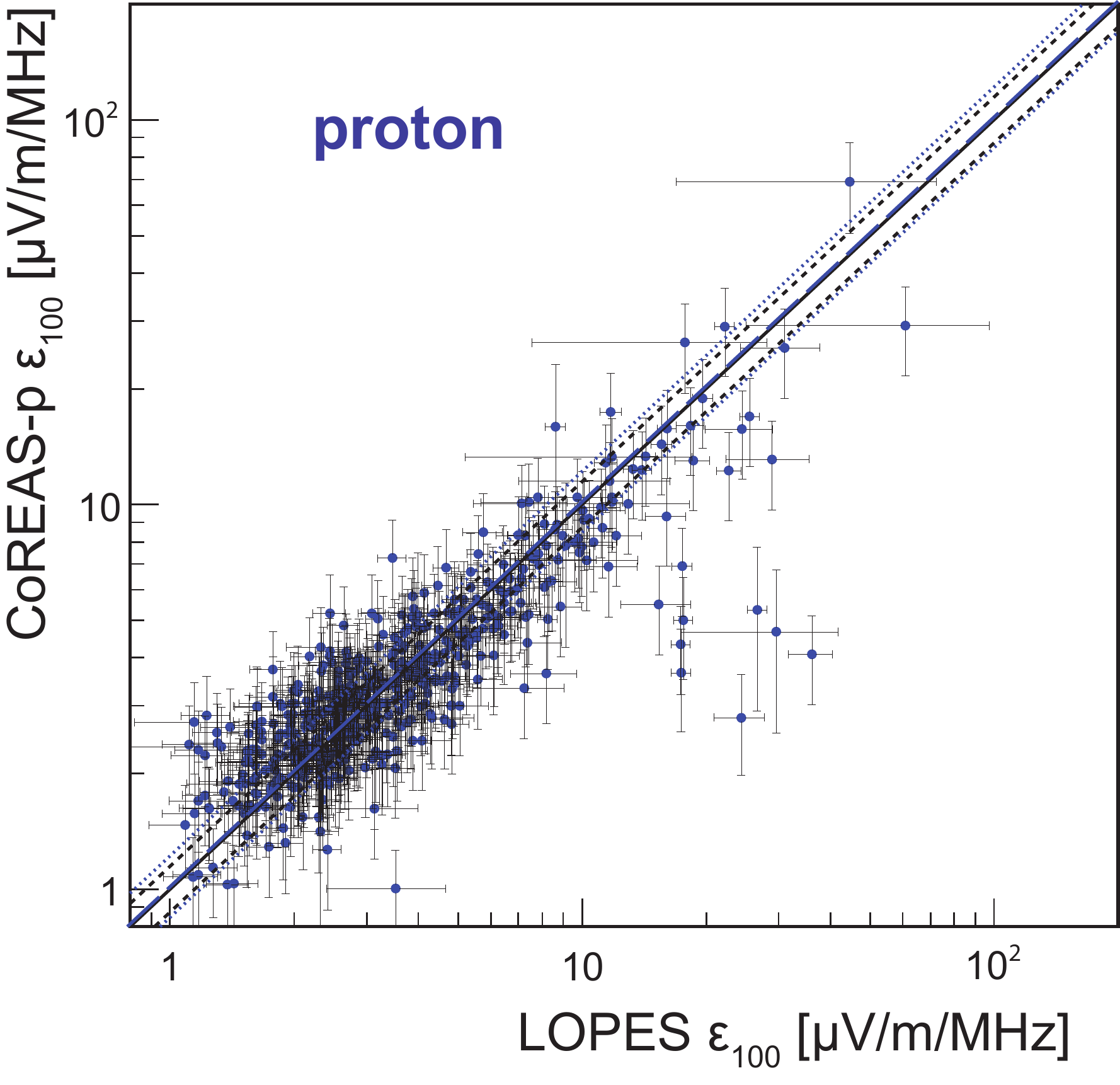}
    \hfill
    \includegraphics[width=0.48\textwidth]{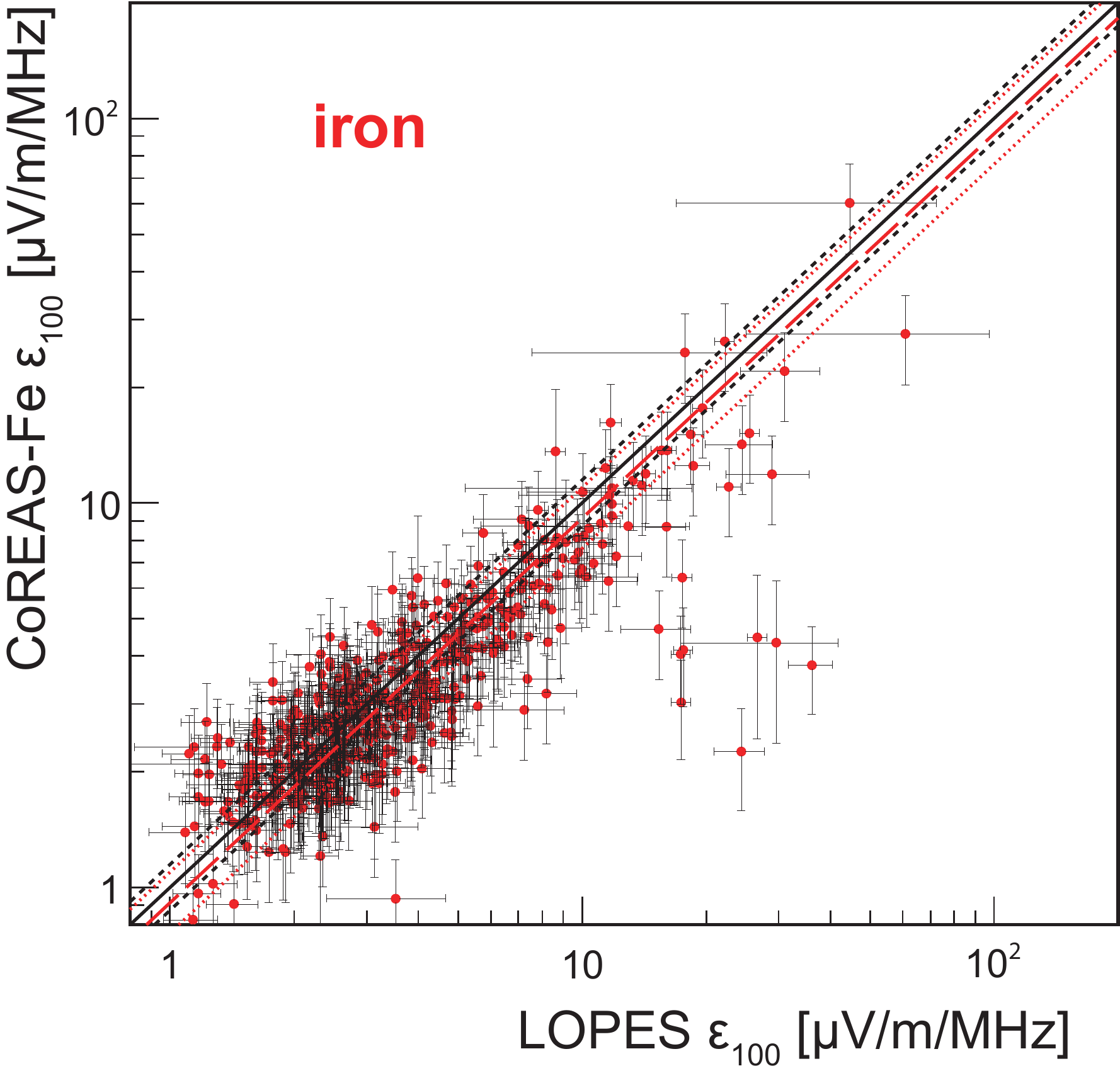}
    \caption{Per-event comparison of $\epsilon_{100}$ derived from 
    LOPES measurements and from CoREAS simulations for a simulation set of 
    proton-induced showers (left) and iron-induced showers (right). 
    The black lines indicate the 1:1 expectation (solid) and the 
    systematic scale uncertainty of the amplitude calibration corresponding to 
    16\% (dashed). The colored lines mark the actual 
    correlation between simulations and data (long-dashed) and the 
    associated systematic uncertainty of the predicted amplitudes of 20\% arising from the energy 
    reconstruction of KASCADE-Grande (dotted).}
    \label{fig:epscomparison}
\end{figure*}

We perform our standard analysis procedure on the LOPES data, which 
identifies the pulses in individual antennas, quantifies their pulse 
amplitudes and fits an exponential lateral distribution to the 
measured radio amplitude $\epsilon$ as a function of axis distance $d$:
\begin{equation}
\epsilon(d) = \epsilon_{100}\,\exp[-\eta(d - 100\,\mathrm{m})]
\end{equation}
The two parameters derived from this fit are the amplitude at 
a distance of 100~m from the shower axis, $\epsilon_{100}$, and the slope 
parameter characterizing the steepness of the lateral distribution, $\eta$.
The parameter $\epsilon_{100}$ is an estimator for the cosmic ray 
energy \cite{ecrs2014}, and the slope parameter $\eta$ can be related to the depth of shower 
maximum \cite{muonprd}. 

In Fig.\ \ref{fig:influenceofcalib} we illustrate the change of 
$\epsilon_{100}$ and the amplitude of the cross-correlation beam 
with respect to the LOPES data presented in \cite{simcomparison}.
On average, the amplitude drops by a factor of $2.6 \pm 0.2$ when 
switching to the revised calibration, where the factor is the same 
within uncertainties for $\epsilon_{100}$ and the amplitude of the cross-correlation beam.
This change is purely due to the new calibration, the minor improvements in 
the analysis pipeline had no significant effect on the amplitude scale. 

\begin{figure*}[t]
    \centering
    \includegraphics[width=0.49\textwidth,clip=true,trim=1cm 7.5cm 1cm 7.5cm]{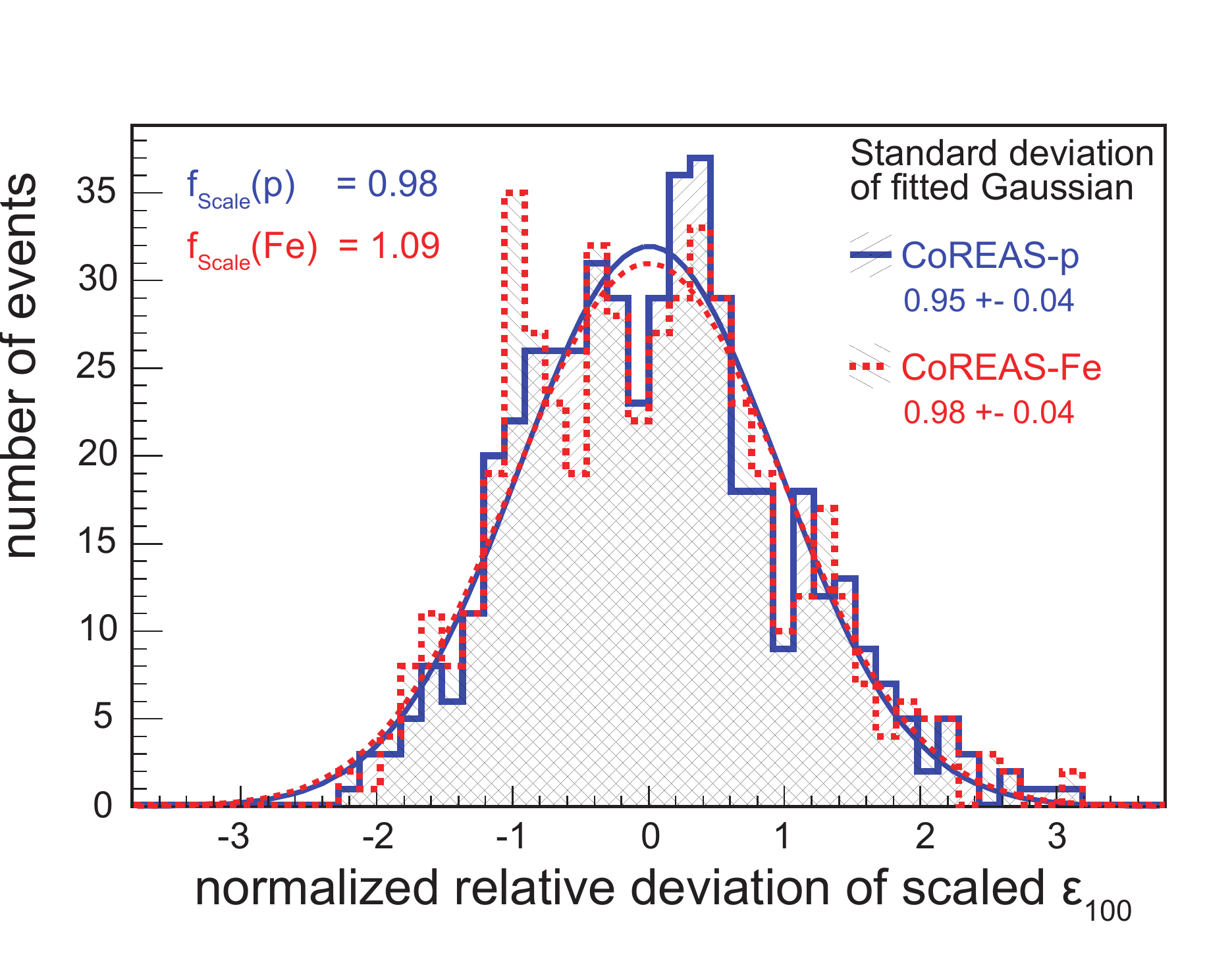}
    \includegraphics[width=0.49\textwidth,clip=true,trim=1cm 7.5cm 1cm 7.5cm]{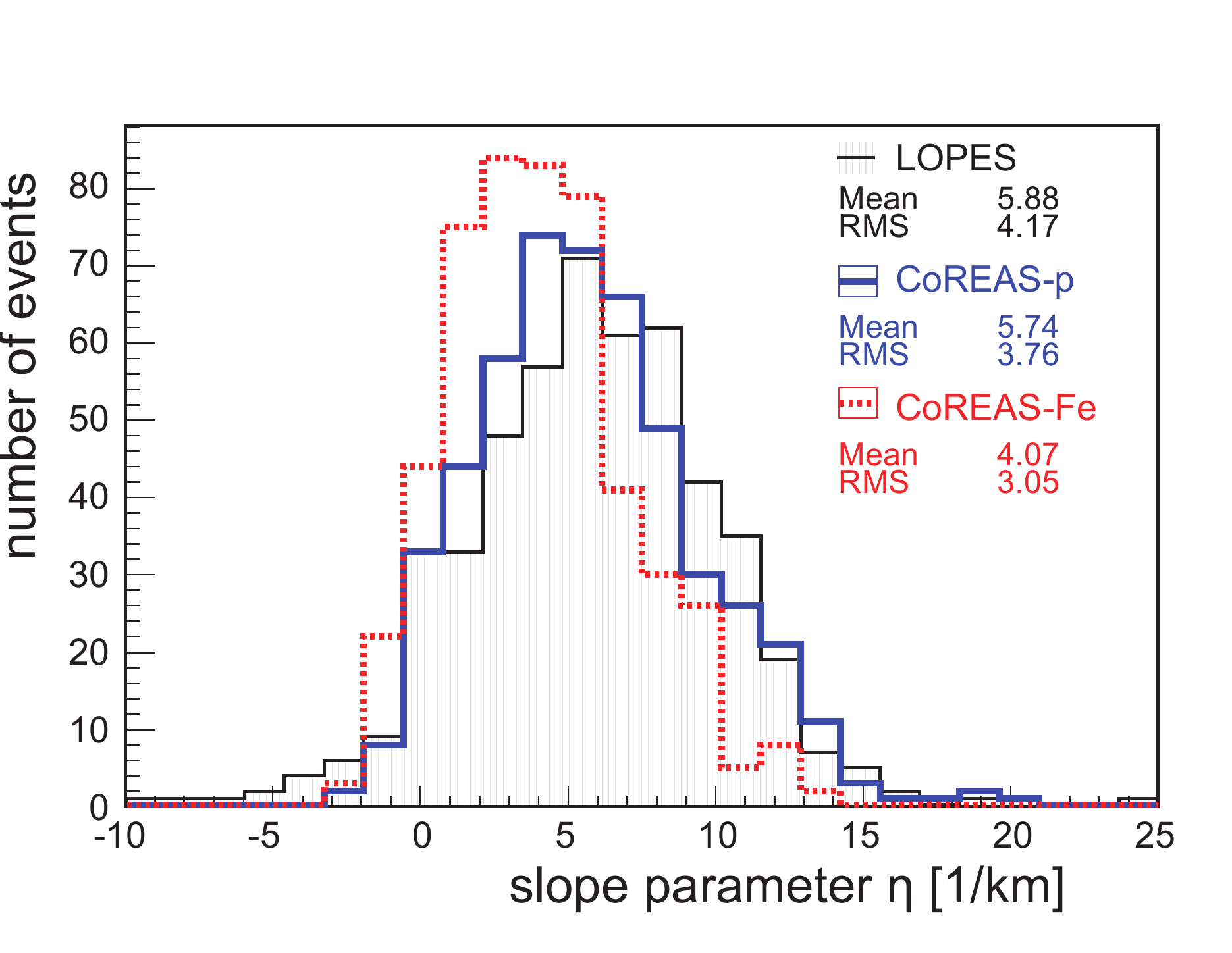}
    \caption{Left: Histogram of the per-event deviations of $\epsilon_{100}$ between LOPES 
    data and CoREAS simulations normalized by the per-event 
    uncertainty in $\epsilon_{100}$. The mean offset factors between 
    simulations and data of 0.98 for proton simulations and 1.09 for 
    iron simulations have been corrected before histogramming (to 
    obtain distributions with a mean of zero). Right: Comparison of the $\eta$ 
    distributions derived from LOPES data and CoREAS simulations.}
    \label{fig:epsetahistos}
\end{figure*}

The slope parameter $\eta$ was affected by both the minor analysis 
improvements and the revised calibration, but changes were within the
method-related systematic uncertainty of approximately 1/km and in fact
compensated each other (see next section).

\section{Comparison of data and CoREAS simulations}

In a next step, we repeat the comparison of LOPES data with CoREAS 
simulations. We perform the same analysis as presented in 
\cite{simcomparison}. This means that for each LOPES event we compare with one CoREAS simulation 
of a proton-induced air shower and one CoREAS simulation of an 
iron-induced air shower. The peak amplitudes for the individual 
simulated antennas are determined from the bandpass-filtered simulations
output. (The influence of a complete detector simulation 
has also been investigated and is presented in another contribution 
at this conference\cite{LOPESoverview}.) The simulated peak amplitudes
are used as input for the same lateral distribution fit 
as applied to the LOPES data. Afterwards, the parameters 
$\epsilon_{100}$ and $\eta$ as derived for LOPES data and CoREAS 
simulations are compared.

In Fig.\ \ref{fig:epscomparison}, we compare $\epsilon_{100}$ as 
measured with LOPES and simulated with CoREAS for each individual 
shower. There is good agreement between the simulated and 
measured data, well within the 16\% systematic uncertainty on the 
absolute amplitude scale of the calibration. Previously, this 
comparison yielded a clear discrepancy between the CoREAS simulations 
and LOPES data \cite{simcomparison}. (The few outliers were already present in the 
previous analysis; the reason for those is unknown.)

A more quantitative view of these data is given 
in Fig.\ \ref{fig:epsetahistos} (left). The histogram shows the 
deviation between the measured and simulated $\epsilon_{100}$ after 
correcting for the mean $\epsilon_{100}$ offset factors of the proton and iron data sets.
The mean offset factor for the proton simulations is 0.98, 
i.e.\ the mean deviation between $\epsilon_{100}$ values predicted by CoREAS and measured by 
LOPES in only 2\% for proton, and 9\% for iron simulations. This agreement is well within the systematic uncertainty 
of the amplitude calibration that the manufacturer of the reference 
source quantifies as 16\%. The width of the distributions correspond to one standard 
deviation, i.e., the distribution of the deviations conforms to the 
expectations from statistical uncertainties.

\begin{figure}[t]
    \centering
    \includegraphics[width=0.55\textwidth]{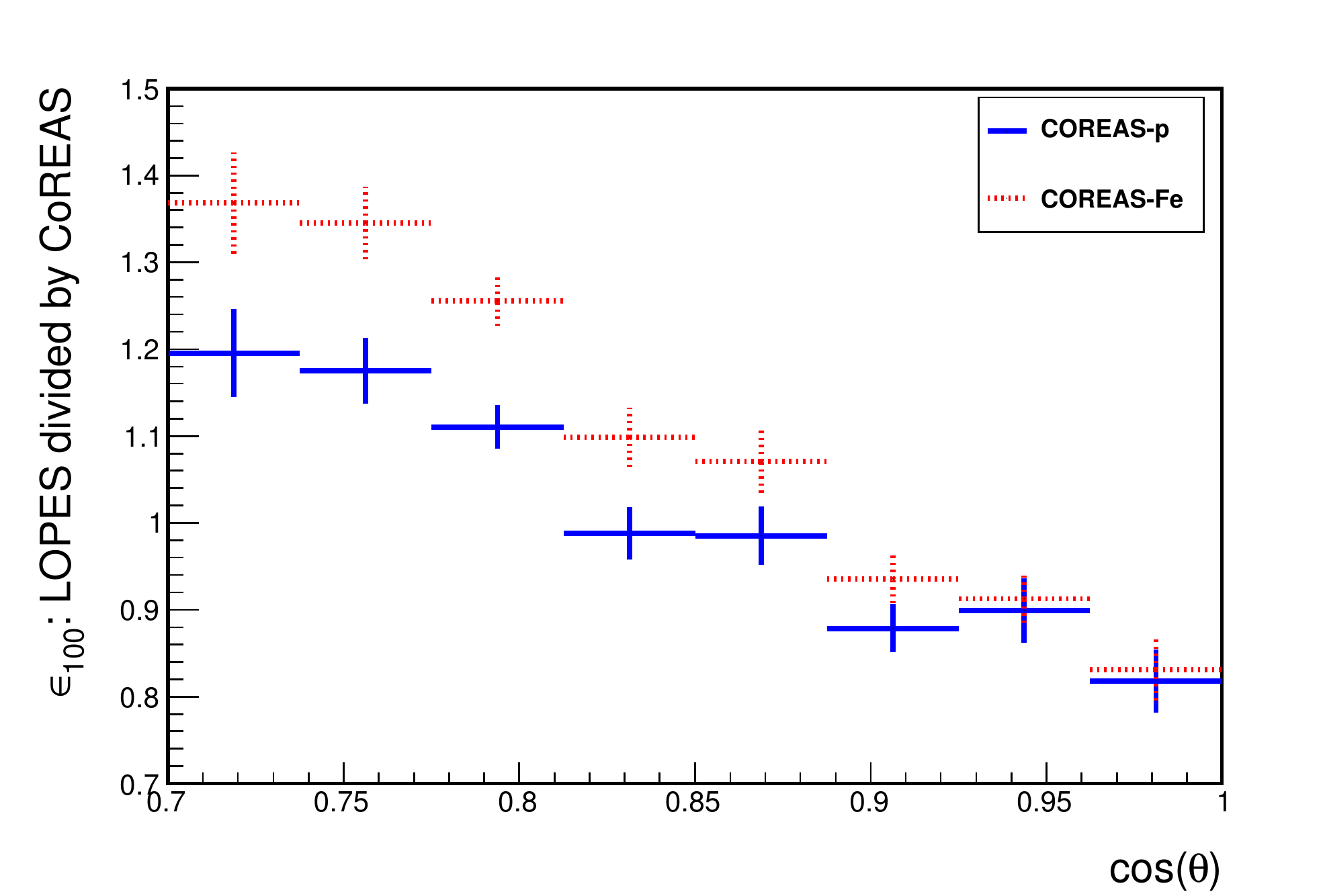}
    \caption{Offset factor for $\epsilon_{100}$ between LOPES data and 
    CoREAS simulations as a function of air shower zenith angle.}
    \label{fig:zenithdependence}
\end{figure}

For the comparison of $\eta$, we do not directly compare individual 
measured showers and simulations, as the depth of shower maximum of 
each shower is unknown but influences $\eta$. Thus, we compare the 
$\eta$-distribution derived from data with those derived from the 
CoREAS simulation sets for proton- and iron-induced showers in Fig.\ 
\ref{fig:epsetahistos} (right). The distributions agree well between 
simulations and data, as was already the case in the previous 
comparison \cite{simcomparison}. The mean of the measured 
distribution is practically unchanged from the previously published 
result, well within the systematic uncertainty of approximately 1/km on $\eta$.

In a recent publication \cite{nunzia} we noted that there is a slight discrepancy in 
the amplitude scaling with zenith angle as predicted by CoREAS 
simulations and LOPES data. We investigated whether this discrepancy 
was removed by the revised calibration or the minor analysis 
improvements. As can be seen in Fig.\ 
\ref{fig:zenithdependence}, the discrepancy remains. The deviation is of the order that can 
be expected from uncertainties in the antenna directivity pattern and 
thus does not necessarily constitute a problem. Nevertheless, future 
analyses should pay attention to the zenith-dependence of the radio 
emission in comparison with predictions from simulation codes.

\section{Influence on published results}

The revised calibration directly influences the absolute amplitudes 
measured with LOPES, which on average turn out to be a factor of $2.6 
\pm 0.2$ lower than those determined with the old calibration. This correction factor can 
be applied to all amplitude values previously published by LOPES, in 
particular any energy correlation results \cite{ecrs2014,nunzia}. 
Our reanalysis demonstrated that no other relevant 
reconstruction quantities are affected significantly by the 
updated recalibration in conjunction with analysis improvements.
In particular, results on the lateral slope $\eta$ remain valid.

\section{Conclusion}

Previously, LOPES had reported a discrepancy between measured 
amplitudes for radio emission from extensive air showers and 
amplitudes predicted by state-of-the-art simulations with the CoREAS simulation 
code. A thorough investigation revealed that unsuitable calibration 
values had been used in the previous calibration of LOPES. New 
calibration values, suitable for \emph{free-space} conditions, have been 
acquired from the manufacturer of the reference source. A reanalysis 
with the new calibration reveals that on average, amplitudes 
measured with LOPES drop by a factor of $2.6 \pm 0.2$ with respect to the old 
calibration. The slope of the lateral distribution of the radio signal 
is not affected significantly by the revised calibration in 
conjunction with minor analysis improvements.

With the new calibration, LOPES data and CoREAS 
simulations are in very good agreement, well within the 16\% systematic 
uncertainty of the absolute amplitude scale of the calibration source.
The only remaining issue is a slight discrepancy between the zenith-angle 
dependence of the electric field amplitudes observed in LOPES data and CoREAS
simulations, which could, however, be explained by the uncertainties of
the antenna directivity pattern. The (obsolete) REAS3.11 simulations
are no longer in agreement with the LOPES data within systematic uncertainties using the revised 
calibration.

More precise tests of the absolute amplitude predictions of state-of-the-art simulation codes require 
experimental data with a smaller systematic scale uncertainty. Direct 
comparisons between Tunka-Rex \cite{hiller}, LOFAR \cite{lofar} and LOPES data, however, will not 
be limited by the 16\% scale uncertainty, as all three experiments 
have been cross-calibrated with the LOPES reference source, including 
a cross-check with Galactic noise \cite{lofarcalib}.

\end{document}